\documentclass[twocolumn,showpacs,showkeys,preprintnumbers,amssymb,aps,superscriptaddress,prb]{revtex4}
\usepackage{graphicx}
\usepackage{dcolumn}
\usepackage{bm}

\begin{document}

\preprint{APS/123-QED}

\title{Exact evidence for the spontaneous antiferromagnetic long-range order 
in the two-dimensional hybrid model of localized Ising spins and itinerant electrons}
\author{Jozef Stre\v{c}ka} 
\affiliation{Department of Theoretical Physics and Astrophysics, 
Faculty of Science, \\ P. J. \v{S}af\'{a}rik University, Park Angelinum 9,
040 01 Ko\v{s}ice, Slovak Republic}
\email{jozef.strecka@upjs.sk, jozkos@pobox.sk} 
\homepage{http://158.197.33.91/~strecka}
\author{Akinori Tanaka}
\affiliation{Department of General Education, Ariake National College of Technology, \\
Omuta, Fukuoka 836-8585, Japan}
\author{Lucia \v{C}anov\'a}
\affiliation{Department of Applied Mathematics, Faculty of Mechanical Engineering, \\ 
Technical University, Letn\'a 9, 042 00 Ko\v{s}ice, Slovak Republic}
\author{Taras Verkholyak}
\affiliation{Institute for Condensed Matter Physics, National Academy of Sciences of Ukraine, \\
        1 Svientsitskii Street, L'viv-11, 79011, Ukraine}

\date{\today}
             
\begin{abstract}
The generalized decoration-iteration transformation is adopted to treat exactly a hybrid model 
of doubly decorated two-dimensional lattices, which have localized Ising spins at their nodal 
lattice sites and itinerant electrons delocalized over pairs of decorating sites. Under 
the assumption of a half filling of each couple of the decorating sites, the investigated 
model system exhibits a remarkable spontaneous antiferromagnetic long-range order with 
an obvious quantum reduction of the staggered magnetization. It is shown that the critical 
temperature of the spontaneously long-range ordered quantum antiferromagnet displays 
an outstanding non-monotonic dependence on a ratio between the kinetic term and 
the Ising-type exchange interaction.
\end{abstract}

\pacs{05.50.+q, 05.70.Jk, 64.60.Cn, 64.60.De, 75.10.-b, 75.10.Hk, 75.10.Jm, 75.30.Kz, 75.40.Cx} 
\keywords{quantum antiferromagnetism, phase transitions and critical phenomena, exact results}

\maketitle

\section{\label{sec:intro} Introduction}

Exactly solvable lattice-statistical models are of appreciable scientific interest as they 
bring a valuable insight into diverse aspects of quantum, cooperative, and critical phenomena.\cite{baxt82,matt93,sach99,lavi99,lieb04,suth04,diep04,wu09} It should be mentioned, 
however, that sophisticated mathematical methods must be usually employed when searching for 
an exact treatment of even relatively simple interacting many-body systems, while an exact treatment 
of more realistic or more complex models is often quite unfeasible or is accompanied with 
a substantial increase of computational difficulties. From this point of view, the exact mapping technique based on generalized algebraic transformations \cite{fish59,syoz72,roja09} belongs to 
the simplest mathematical methods, which allows to obtain the exact solution of more complicated model from a precise mapping relationship with a simpler exactly solved model. Following Fisher's 
ideas,\cite{fish59} an arbitrary statistical-mechanical system (even of quantum nature) that 
merely interacts with either two or three outer Ising spins may be in principle replaced by the effective interactions between the outer Ising spins through appropriately chosen decoration-iteration \cite{syoz51} or star-triangle \cite{onsa44} mapping transformations. Even although the concept based on the generalized algebraic transformations has been worked out by Fisher more than a half century ago,\cite{fish59} the algebraic mapping transformations were initially widely used to treat only lattice-statistical models consisting of the Ising spins (see Ref. \onlinecite{syoz72} and 
references cited therein) before this conceptually simple approach was finally applied 
to hybrid models composed of the Ising and classical Heisenberg spins,\cite{gonc82,hori83,gonc84,gonc86,sant86,sant95} as well as, the Ising 
and quantum Heisenberg spins.\cite{stre02,jasc02,stre04,stre06,stre08,yao08,valv09,stre09}  
   
Another interesting application of the algebraic mapping transformations has recently been suggested 
by Pereira \textit{et al}. \cite{pere08,pere09} when applying the generalized decoration-iteration transformation to an intriguing diamond-chain model of interacting spin-electron system. In this diamond-chain model, the nodal lattice sites are occupied by the localized Ising spins and 
the mobile electrons can freely move on a couple of interstitial decorating sites symmetrically 
placed in between two localized Ising spins. The main aim of this work is to treat exactly 
an analogous two-dimensional (2D) hybrid model defined on doubly decorated planar lattices, which 
should provide a deeper insight into how itinerant character of the mobile electrons will influence phase transitions and critical phenomena of this interacting spin-electron system.   

The outline of this paper is as follows. In Section \ref{sec:model}, we first provide a rather 
detailed description of the model under investigation together with the most crucial steps of the 
exact mapping method, which enables us to obtain exact closed-form expressions for the critical temperature, order parameter and other relevant thermodynamic quantities. The most interesting 
results for the ground state, the finite-temperature phase diagram and thermodynamics are 
presented and detailed discussed in Section \ref{sec:result}. Finally, the summary of 
the most important scientific achievement is mentioned with several concluding remarks 
in Section \ref{sec:conc}. 

\section{\label{sec:model} Model and method}

\begin{figure}[t]
\includegraphics[width=8cm]{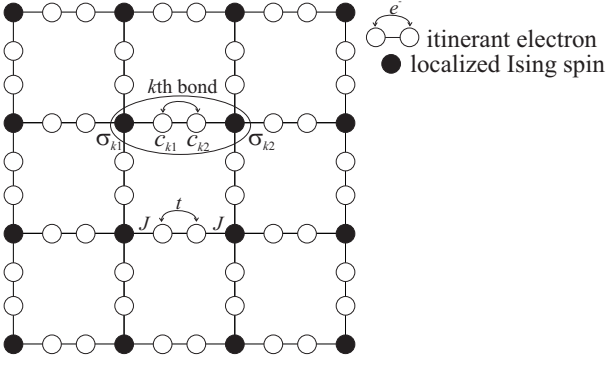}
\caption{A part of the doubly decorated square lattice, which has one localized Ising spin 
at each nodal lattice site (full circle) and two mobile electrons at each couple of decorating 
sites (empty circles). The ellipse demarcates the localized Ising spins and itinerant electrons described through the $k$th bond Hamiltonian (\ref{ham}).}
\label{fig1}
\end{figure}

Let us consider a hybrid lattice-statistical model of interacting spin-electron system on 
doubly decorated 2D lattices, which have one localized Ising spin at each nodal lattice site 
and two delocalized mobile electrons at each couple of decorating sites. The magnetic 
structure of the model under investigation is schematically depicted in figure \ref{fig1} 
on the particular example of the doubly decorated square lattice.  
For further convenience, the total Hamiltonian can be defined as a sum over bond Hamiltonians 
$\hat{\cal H} = \sum_k \hat{\cal H}_k$, where each bond Hamiltonian $\hat{\cal H}_k$ involves 
all the interaction terms of two itinerant electrons from the $k$th bond of the doubly decorated 
2D lattice
\begin{eqnarray}
\hat{\cal H}_k = \! \! \! &-& \! \! \! t \left( c^{\dagger}_{k1, \uparrow} c^{}_{k2, \uparrow} 
                          + c^{\dagger}_{k1, \downarrow} c^{}_{k2, \downarrow}
                          + c^{\dagger}_{k2, \uparrow} c^{}_{k1, \uparrow}  
                          + c^{\dagger}_{k2, \downarrow} c^{}_{k1, \downarrow} \right)	\nonumber \\
                 \! \! \! &-& \! \! \! 
          \frac{J}{2} \hat{\sigma}_{k1}^z \left( c^{\dagger}_{k1, \uparrow} c^{}_{k1, \uparrow} 
                                        -  c^{\dagger}_{k1, \downarrow} c^{}_{k1, \downarrow} \right) 
                                        \nonumber \\
              \! \! \! &-& \! \! \! 
          \frac{J}{2} \hat{\sigma}_{k2}^z \left( c^{\dagger}_{k2, \uparrow} c^{}_{k2, \uparrow} 
                                -  c^{\dagger}_{k2, \downarrow} c^{}_{k2, \downarrow} \right). \label{ham}
\end{eqnarray}
Here, $c^{\dagger}_{k \alpha, \gamma}$ and $c^{}_{k \alpha, \gamma}$ ($\alpha = 1,2$, 
$\gamma = \uparrow, \downarrow$) denote usual creation and annihilation fermionic operators 
and $\hat{\sigma}_{k \alpha}^z$ is the standard spin-1/2 operator with the eigenvalues 
${\sigma}_{k \alpha}^z = \pm 1/2$. The hopping parameter $t$ takes into account kinetic energy 
of the mobile electrons and the exchange integral $J$ describes the Ising-type interaction 
between the itinerant electrons and their nearest Ising neighbors. 

A crucial step of our approach lies in the calculation of the partition function. Owing to a validity of the commutation relation between different bond Hamiltonians $[\hat{\cal H}_i, \hat{\cal H}_j] = 0$, the total partition function ${\cal Z}$ can be partially factorized into a product of the bond partition functions ${\cal Z}_k$
\begin{eqnarray}
{\cal Z} = \displaystyle \sum_{\{ \sigma_i \}} \prod_{k=1}^{Nq/2} 
           \mbox{Tr}_{k} \exp \left(- \beta \hat{\cal H}_k \right) 
          = \displaystyle \sum_{\{ \sigma_i \}} \prod_{k=1}^{Nq/2} {\cal Z}_k, 
\label{pf}
\end{eqnarray}
where $\beta = 1/(k_{\rm B} T)$, $k_{\rm B}$ is Boltzmann's constant, $T$ is the absolute temperature,
$N$ is the total number of the Ising spins (i.e. the nodal lattice sites) and $q$ is their coordination number (i.e. the number of nearest-neighbor decorating sites). Next, the symbol $\sum_{\{\sigma_i \}}$ denotes a summation over all possible spin configurations of the localized Ising spins and the symbol $\mbox{Tr}_{k}$ stands for a trace over degrees of freedom of two mobile electrons 
from the $k$th couple of decorating sites. An explicit form of the bond partition function ${\cal Z}_k$ can be subsequently acquired by a direct diagonalization of the bond Hamiltonian (\ref{ham}). 
The matrix representation of the bond Hamiltonian $\hat{\cal H}_k$ in the orthonormal basis of states $| \psi_i \rangle = \{c^{\dagger}_{k1, \uparrow} c^{\dagger}_{k2, \uparrow}| 0 \rangle$, $c^{\dagger}_{k1, \downarrow} c^{\dagger}_{k2, \downarrow} | 0 \rangle$, $c^{\dagger}_{k1, \uparrow} c^{\dagger}_{k2, \downarrow} | 0 \rangle$, $c^{\dagger}_{k1, \downarrow} c^{\dagger}_{k2, \uparrow} | 0 \rangle$, $c^{\dagger}_{k1, \uparrow} c^{\dagger}_{k1, \downarrow}| 0 \rangle$, $c^{\dagger}_{k2, \uparrow} c^{\dagger}_{k2, \downarrow}| 0 \rangle \}$ ($| 0 \rangle$ labels the empty state) reads
\begin{widetext} 
\begin{eqnarray} 
\langle \psi_j | \hat{{\cal{H}}}_{k}| \psi_i \rangle = \left( 
   \begin{array}{cccccc}
     -h_{k1} - h_{k2} &   0   &   0   &   0   &   0   &   0   \\
      0   &  h_{k1} + h_{k2}  &   0   &   0   &   0   &   0   \\
      0   &   0   &  -h_{k1} + h_{k2} &   0   &  -t   &  -t   \\
      0   &   0   &   0   &  h_{k1} - h_{k2}  &   t   &   t   \\
      0   &   0   &  -t   &   t   &   0   &   0               \\
      0   &   0   &  -t   &   t   &   0   &   0               \\
   \end{array}
\right), 
\label{matrix}
\end{eqnarray}
\end{widetext}
where we have defined two new parameters $h_{k1} = J \sigma_{k1}^z/2$ and $h_{k2} = J \sigma_{k1}^z/2$ that include the Ising-type interaction between the mobile electrons and their nearest-neighbor localized Ising spins. It is noteworthy that the parameters $h_{k1}$ and $h_{k2}$ can alternatively be viewed also as generally non-uniform effective field, which is produced by the localized Ising spins 
on the itinerant electrons situated at the nearest-neighbor decorating sites. The straightforward diagonalization of the Hamiltonian matrix (\ref{matrix}) yields six eigenenergies 
\begin{eqnarray} 
E_{k1,k2} \! \! \! &=& \! \! \! \pm (h_{k1} + h_{k2}), \qquad E_{k3,k4} = 0, \nonumber \\
E_{k5,k6} \! \! \! &=& \! \! \! \pm \sqrt{\left(h_{k1} - h_{k2} \right)^2 + 4 t^2},
\label{eigen}
\end{eqnarray}  
which can be further used for the relevant calculation of the bond partition function ${\cal Z}_k$. 
After tracing out the degrees of freedom of the itinerant electrons, the bond partition function ${\cal Z}_k$ merely depends on spin states of two localized Ising spins and, besides, its explicit form immediately implies a possibility of performing the generalized decoration-iteration transformation \cite{fish59,syoz72,roja09}
\begin{eqnarray}
\! \! \! \! \! \! \! && \! \! \! \! \! {\cal Z}_k = \sum_{i=1}^6 \exp \left( - \beta E_{ki} \right) =  
 2 + 2 \cosh \! \! \left[ \frac{\beta J}{2} (\sigma_{k1}^z + \sigma_{k2}^z) \right] + \nonumber \\  
\! \! \! \! \! \! \! && \! \! \! \! \! 
  2 \cosh \! \! \left[ \frac{\beta}{2} \sqrt{J^2 (\sigma_{k1}^z - \sigma_{k2}^z)^2 + (4t)^2} \right] 
   = A \exp(\beta R \sigma_{k1}^z \sigma_{k2}^z). \nonumber \\ 
\label{dit}
\end{eqnarray} 
The physical meaning of the mapping transformation (\ref{dit}) is to replace the bond partition function ${\cal Z}_k$ by the equivalent expression, which would contain the effective pair interaction $R$ between the localized Ising spins only. The mapping parameters $A$ and $R$ are unambiguously 
given by the 'self-consistency' condition of the algebraic transformation (\ref{dit}), which must 
hold for any combination of spin states of two Ising spins $\sigma_{k1}^z$ and $\sigma_{k2}^z$ involved therein. It can be readily proved that the decoration-iteration transformation (\ref{dit}) indeed represents a set of two independent equations, which directly determine the mapping 
parameters $A$ and $R$   
\begin{eqnarray}
A = (V_1 V_2)^{1/2}, \qquad \beta R = 2 \ln (V_1/V_2),   
\label{mp}       
\end{eqnarray}
that are for the sake of brevity expressed in terms of the newly defined functions $V_1$ and $V_2$            
\begin{eqnarray}
V_1 \!\!\! &=& \!\!\! 2 + 2 \cosh (\beta J/2) + 2 \cosh (2 \beta t),  \nonumber \\
V_2 \!\!\! &=& \!\!\! 4 + 2 \cosh \left[\beta \sqrt{J^2 + (4t)^2}/2 \right].   
\label{fun}            
\end{eqnarray}
Substituting the algebraic transformation (\ref{dit}) into the factorized form of the partition function (\ref{pf}), which physically corresponds to performing the decoration-iteration mapping 
at each bond of the doubly decorated 2D lattice, consequently leads to a simple mapping relationship between the partition function ${\cal Z}$ of the interacting spin-electron system on the doubly decorated 2D lattice and, respectively, the partition function ${\cal Z}_{{\rm IM}}$ of the 
simple spin-1/2 Ising model on the corresponding undecorated lattice with the effective (temperature-dependent) nearest-neighbor interaction $R$
\begin{eqnarray}
{\cal Z} (\beta, J, t) = A^{Nq/2} {\cal Z}_{{\rm IM}} (\beta, R). 
\label{mr}
\end{eqnarray} 
The mapping relation (\ref{mr}) essentially completes our exact calculation of the partition function ${\cal Z}$, since the partition function of the nearest-neighbor spin-1/2 Ising model has been precisely calculated for several 2D lattices (for reviews see Refs.~\onlinecite{syoz72,domb60,mcoy73}). For brevity, let us therefore merely quote the respective results for the partition functions of the spin-1/2 Ising model on the square lattice \cite{onsa44}
\begin{eqnarray}
\lim_{N \to \infty} \!\!\!\!\! && \!\!\!\!\! \frac{1}{N} \ln {\cal Z}_{{\rm IM}} = \ln 2 
+ \frac{1}{4 \pi^2} \int_0^{2 \pi} \!\!\! \int_0^{2 \pi} \!\!\! \ln [\cosh^2 \left(\beta R/2 \right) \nonumber \\
\!\!\! &-& \!\!\! \sinh \left(\beta R/2 \right) \left(\cos \theta + \cos \phi \right)] 
{\rm d} \theta {\rm d} \phi 
\label{pfsquare}
\end{eqnarray}
and the honeycomb lattice \cite{hout50}
\begin{eqnarray}
\!\!\!\! && \!\!\!\! \lim_{N \to \infty} \!\! \frac{1}{N} \ln {\cal Z}_{{\rm IM}} = \ln 2 
+ \frac{1}{16 \pi^2} \int_0^{2 \pi} \!\!\! \int_0^{2 \pi} \!\!\! \ln [\{1 + 
\cosh^3 \! \left(\beta R/2 \right) \nonumber \\
\!\!\!\! && \!\!\!\! -
\sinh^2 \left(\beta R/2 \right) \left[\cos \theta + \cos \phi + \cos(\theta + \phi) \right]\}/2] 
{\rm d} \theta {\rm d} \phi. 
\label{pfhoney}
\end{eqnarray}
In what follows, the precise mapping relationship (\ref{mr}) between the partition functions 
will be used as a starting point for performing a rather comprehensive analysis of the critical behavior, the order parameter, and basic thermodynamic quantities.

\subsection{Critical condition}
In order to locate a critical point of the investigated spin-electron system, one may take advantage 
of the fact that the partition function always becomes non-analytic at a critical point. It can 
be easily understood from Eqs.~(\ref{mp})--(\ref{mr}) that the mapping parameter $A$ cannot cause 
a non-analytic behavior of the partition function ${\cal Z}$ and thus, the investigated spin-electron system becomes critical if and only if the corresponding spin-1/2 Ising model with the effective coupling $\beta R$ becomes critical as well. Accordingly, the critical points of the investigated spin-electron system can be straightforwardly obtained from a comparison of the effective temperature-dependent coupling $\beta R$ with the relevant 
critical point of the corresponding spin-1/2 Ising model on the undecorated lattice. However, 
the mathematical structure of the mapping parameter $R$ has another important consequences on a critical behavior. More specifically, one may easily prove from Eq. (\ref{fun}) a general validity of the inequality $V_1<V_2$, which in compliance with the definition (\ref{mp}) implies the antiferromagnetic nature of the effective interaction ($R<0$) in the spin-1/2 Ising model on the corresponding undecorated lattice. Owing to this fact, the interacting spin-electron system is always mapped on the spin-1/2 Ising model with the antiferromagnetic nearest-neighbor interaction, which has a non-zero critical temperature only on the 2D loose-packed lattices such as square \cite{onsa44} and honeycomb \cite{hout50,temp50,newe50,husi50,syoz50} lattices. As a matter of fact, 
it is well known dictum that the spin-1/2 Ising model with the antiferromagnetic nearest-neighbor interaction on close-packed lattices like triangular \cite{hout50,syoz50,wann50} and kagom\'e \cite{syoz51,kano53} lattices does not exhibit a spontaneous long-range order at any finite temperature 
and consequently, it does not have any finite-temperature critical point that would correspond 
to the order-disorder transition. 

Bearing all this in mind, we will consider hereafter only the interacting spin-electron system 
on the doubly decorated 2D loose-packed lattices. It is worthwhile to remark that the critical temperature of the antiferromagnetic spin-1/2 Ising model on the loose-packed lattices is equal to 
the one of the ferromagnetic model and hence, the critical points for the spin-electron system on the doubly decorated square and honeycomb lattices readily follow from the conditions 
$\beta_{\rm c} |R| = 2 \ln (1 + \sqrt{2})$\cite{onsa44} and, respectively, 
$\beta_{\rm c} |R| = 2 \ln(2 + \sqrt{3})$,\cite{hout50} where 
$\beta_{\rm c} = 1/(k_{\rm B} T_{\rm c})$ and $T_{\rm c}$ is the critical temperature. 

\subsection{Staggered magnetization}

As a direct consequence of the mapping equivalence with the antiferromagnetic spin-1/2 Ising model 
on the relevant undecorated loose-packed lattice, one should anticipate predominantly 
antiferromagnetic character of the interacting spin-electron system on the doubly decorated 
2D loose-packed lattice as well. Let us therefore calculate the staggered magnetization as 
the most common order parameter inherent to the antiferromagnetic spin alignment. Using the exact mapping theorems developed by Barry \textit{et al}.,\cite{barr88,khat90,barr91,barr95} the spontaneous 
staggered magnetization of the localized Ising spins can easily be calculated from the 
exact spin identity
\begin{eqnarray}
m_i \equiv \frac{1}{2} \langle \hat{\sigma}_{k1}^z - \hat{\sigma}_{k2}^z \rangle_{t,J} 
= \frac{1}{2} \langle \hat{\sigma}_{k1}^z - \hat{\sigma}_{k2}^z \rangle_{R} 
\equiv m_{\rm IM} (\beta R), 
\label{mi}       
\end{eqnarray} 
where the symbols $\langle \ldots \rangle_{t,J}$ and $\langle \ldots \rangle_{R}$ denote 
standard canonical ensemble average performed within the interacting spin-electron model on the doubly 
decorated 2D lattice and, respectively, its equivalent spin-1/2 Ising model on the corresponding undecorated 2D lattice.\cite{acom} The exact spin identity (\ref{mi}) furnishes an accurate proof that the staggered magnetization of the Ising sublattice in the interacting spin-electron model on the doubly decorated 2D lattice directly equals the staggered magnetization of the spin-1/2 Ising model on the corresponding undecorated 2D lattice with the antiferromagnetic nearest-neighbor interaction $R<0$. However, the spontaneous staggered magnetization of the antiferromagnetic spin-1/2 Ising model 
on the loose-packed 2D lattices precisely coincides with the spontaneous magnetization of the ferromagnetic model, i.e. the quantity, which has exactly been determined for several 2D Ising models 
(see Ref. \onlinecite{lin92} and references cited therein). In this regard, the staggered magnetization 
of the antiferromagnetic spin-1/2 Ising model on the square \cite{yang52} and honeycomb \cite{naya54} lattices for instance read
\begin{eqnarray}
m_{\rm IM} \! \! \! &=& \! \! \! \frac{1}{2} \left[ 1 - \frac{16 x^4}{(1-x^2)^4} \right]^{1/8}
\! \! \! \! \! \!, 
\qquad {\rm (square)} \label{m0}   \\
m_{\rm IM} \! \! \! &=& \! \! \! 
    \frac{1}{2} \left[ 1 - \frac{16 x^3 (1 + x^3)}{(1-x)^3(1-x^2)^3} \right]^{1/8}\! \! \! \! \! \!, 
\qquad {\rm (honeycomb)} \nonumber     
\end{eqnarray} 
where $x = \exp(- \beta R/2)$. The above formulas complete our calculation of the staggered magnetization of the localized Ising spins when substituting the exact expression (\ref{m0}) 
with the appropriately chosen effective coupling (\ref{mp})--(\ref{fun}) 
into the exact spin identity (\ref{mi}). 

On the other hand, the staggered magnetization of the itinerant electrons delocalized over pairs 
of decorating sites can be calculated with the aid of the generalized Callen-Suzuki identity \cite{call63,suzu65,saba81,saba85,balc02}
\begin{eqnarray}
\langle f (c^{\dagger}_{k \alpha, \gamma}, c^{}_{k \alpha, \gamma}) \rangle_{t,J}
  \! = \left \langle 
  \frac{\mbox{Tr}_k f (c^{\dagger}_{k \alpha, \gamma}, c^{}_{k \alpha, \gamma})  
        \exp(- \beta \hat{\cal H}_k)}{\mbox{Tr}_k \exp(- \beta \hat{\cal H}_k)} 
        \right \rangle_{\! \! t,J}\! \! \! \! \! \!,  \nonumber \\
\label{csi}
\end{eqnarray} 
where $\alpha=1,2$, $\gamma = \uparrow,\downarrow$, and $f$ is in principle an arbitrary function of the creation and annihilation fermionic operators from the $k$th bond Hamiltonian (\ref{ham}). With the help of the exact identity (\ref{csi}), the spontaneous staggered magnetization of 
the itinerant electrons can be calculated from the expression 
\begin{eqnarray}
m_e = \frac{1}{2} \langle (\hat{S}_{k1}^z - \hat{S}_{k2}^z) \rangle_{t,J}
 \!  = \left \langle \frac{1}{4 \beta {\cal Z}_k} \left( \frac{\partial {\cal Z}_k}{\partial h_{k1}} - 
                    \frac{\partial {\cal Z}_k}{\partial h_{k2}} \right)
                                        \right \rangle_{t,J}\! \! \! \! \! \!,  \nonumber \\
\label{me}
\end{eqnarray} 
where $\hat{S}_{k \alpha}^z = (c^{\dagger}_{k \alpha, \uparrow} c^{}_{k \alpha, \uparrow} -  c^{\dagger}_{k \alpha, \downarrow} c^{}_{k \alpha, \downarrow})/2$ marks the $z$th component 
of the spin operator of the mobile electron. After a straightforward calculation based 
on the differential operator technique $\exp(a \partial/\partial x + b \partial/\partial y) g(x,y) = g(x+a, y+b)$ \cite{honm79,kane93} and the exact van der Waerden identity $\exp(c \sigma^z) 
= \cosh(c/2) + 2 \sigma^z \sinh(c/2)$, the staggered magnetization of the itinerant electrons can be related to the staggered magnetization of the localized Ising spins through the precise formula
\begin{eqnarray}
m_e = m_i \frac{J}{\sqrt{J^2 + (4t)^2}} 
\frac{\sinh \left[\frac{\beta}{2} \sqrt{J^2 + (4t)^2} \right]}
     {2 + \cosh \left[\frac{\beta}{2} \sqrt{J^2 + (4t)^2} \right]}.  
\label{mag}
\end{eqnarray}
Since the exact expression for the staggered magnetization of the localized Ising spins is already known from Eqs. (\ref{mi})--(\ref{m0}), the formula (\ref{mag}) provides the relevant exact result
for the staggered magnetization of the itinerant electrons hopping between the decorating sites.

\subsection{Thermodynamics} 

Before concluding this section, it is worthy of notice that several basic thermodynamic quantities 
can also be easily derived from the exact mapping equivalence (\ref{mr}) between the partition functions ${\cal Z}$ and ${\cal Z}_{\rm IM}$. For instance, the Helmholtz free energy $F$, 
the internal energy $U$, the entropy $S$, and the specific heat $C$, can directly be computed 
from the basic relations of thermodynamics and statistical physics such as 
\begin{eqnarray}
F = - k_{\rm B} T \ln {\cal Z}, \, \, \, U = - \frac{\partial \ln {\cal Z}}{\partial \beta}, \, \, \,
S = - \frac{\partial F}{\partial T}, \,\, \, C = \frac{\partial U}{\partial T}. \nonumber
\label{tsp}
\end{eqnarray}  

\section{\label{sec:result} Results and discussion}

Now, let us proceed to a discussion of the most interesting results obtained for the interacting spin-electron system on the doubly decorated 2D lattices. Before doing this, however, the relations (\ref{mp})--(\ref{fun}) might serve in evidence that the effective nearest-neighbor interaction $R$ 
of the spin-1/2 Ising model on the corresponding undecorated lattice is invariant under the transformation $J \to -J$. A change of the ferromagnetic Ising interaction $J>0$ to the antiferromagnetic one $J<0$ actually causes only a rather trivial change of the mutual spin orientation 
of the itinerant electrons and their nearest Ising neighbors. This observation would suggest 
that the critical temperature as well as other thermodynamic quantities remain unchanged under 
the sign change $J \to -J$ and thus, one may further consider the ferromagnetic interaction 
$J>0$ without loss of the generality.  

\subsection{Ground state} 

\begin{figure}[t]
\begin{minipage}{8cm}
\includegraphics[width=8cm]{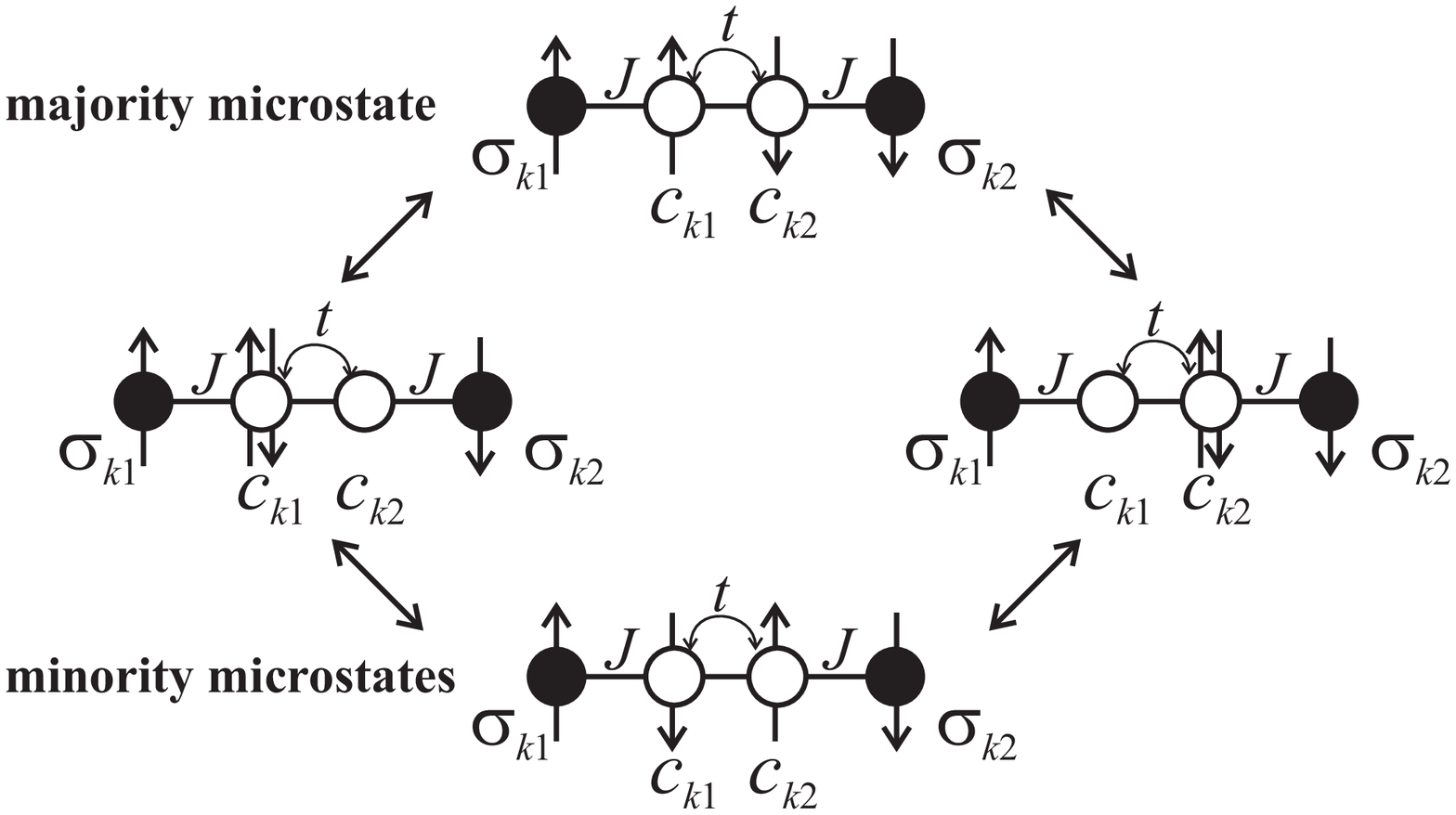}
\vspace{-0.5cm}
\caption{\label{fig2}A schematic representation of the quantum entanglement of two antiferromagnetic  and two non-magnetic ionic states, which occurs in $|{\rm AF} \rangle$ on account of a virtual hopping 
process of the mobile electrons.}
\end{minipage}
\begin{minipage}{8cm}
\vspace{0.2cm}
\includegraphics[width=8cm]{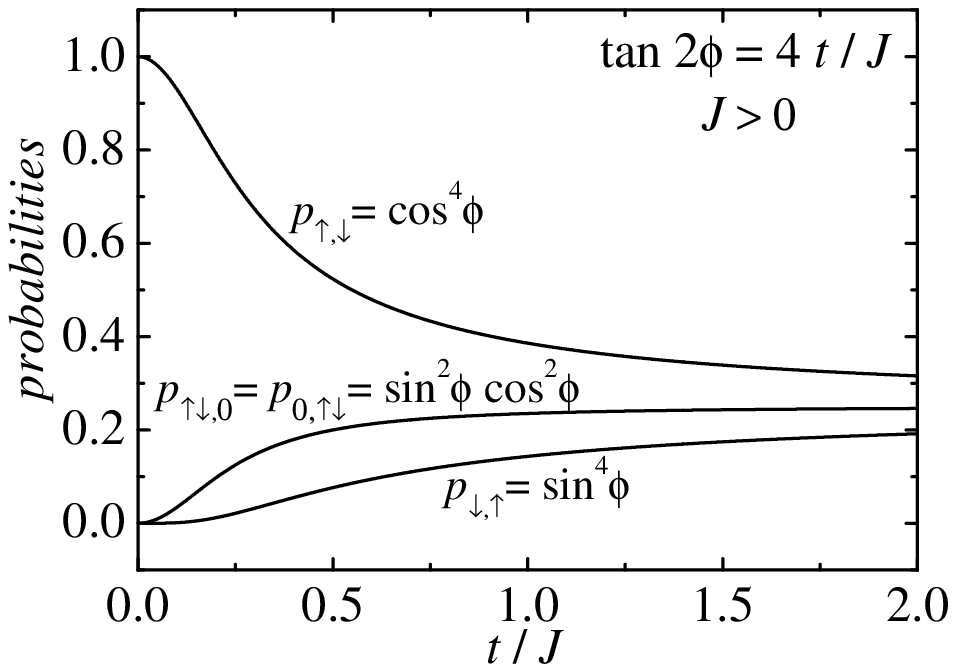}
\vspace{-1.1cm}
\caption{\label{fig3}The probability distribution for the four entangled microstates 
as a function of the ratio between the hopping term $t$ and the Ising exchange constant $J$.}
\end{minipage} 
\end{figure}

First, let us take a closer look at the ground-state behavior. It is quite evident that the ground state will correspond to the lowest-energy eigenvalue of the bond Hamiltonian (\ref{ham}), which can 
be obtained from the eigenenergies (\ref{eigen}) by considering four available configurations 
of the Ising spins $\sigma_{k1}$ and $\sigma_{k2}$ explicitly involved therein. It turns out that 
the lowest-energy eigenstate constitutes a peculiar four-sublattice quantum antiferromagnet, 
which can be characterized through the eigenvector 
\begin{eqnarray}
|{\rm AF} \rangle = \displaystyle \prod_{k=1}^{Nq/2} \! \! \! \! \! \! && \! \! \! \! \! \!
|\!\! \uparrow, \downarrow \rangle_{\sigma_{k1}, \sigma_{k2}} \Biggl[ \frac{1}{2} \left(1 + \frac{J}{\sqrt{J^2 + (4t)^2}} \right) c^{\dagger}_{k1, \uparrow} c^{\dagger}_{k2, \downarrow} 
\nonumber \\ 
\! \! \! &-& \! \! \! \frac{1}{2} \left(1 - \frac{J}{\sqrt{J^2 + (4t)^2}} \right) 
c^{\dagger}_{k1, \downarrow} c^{\dagger}_{k2, \uparrow} \label{gs}  \\
\! \! \! &+& \! \! \! \frac{2t}{\sqrt{J^2 + (4t)^2}} \left(c^{\dagger}_{k1, \uparrow} 
c^{\dagger}_{k1, \downarrow} + c^{\dagger}_{k2, \uparrow} c^{\dagger}_{k2, \downarrow}
\right) \Biggr] | 0 \rangle,
\nonumber
\end{eqnarray}
where the product runs over all bonds of the doubly decorated 2D lattice, the former ket vector determines spin states of the localized Ising spins, and the latter one spin states of mobile electrons. Interestingly, one may also find a much simpler goniometric representation of the lowest-energy eigenstate $|{\rm AF} \rangle$ by introducing the mixing angle $\phi$ through 
the definition $\tan 2 \phi = 4t/J$, which yields for the particular case with $J>0$ \cite{note}
\begin{eqnarray}
|{\rm AF} \rangle \! \! \! &=& \! \! \! \displaystyle \prod_{k=1}^{Nq/2} 
|\!\! \uparrow, \downarrow \rangle_{\sigma_{k1}, \sigma_{k2}} 
\Bigl[ \cos^2 \! \phi \, c^{\dagger}_{k1, \uparrow} c^{\dagger}_{k2, \downarrow} - 
\sin^2 \! \phi \, c^{\dagger}_{k1, \downarrow} c^{\dagger}_{k2, \uparrow}  \nonumber \\
\! \! \! &+& \! \! \! \sin \phi \cos \phi \left( c^{\dagger}_{k1, \uparrow} 
c^{\dagger}_{k1, \downarrow} + c^{\dagger}_{k2, \uparrow} c^{\dagger}_{k2, \downarrow} \right) 
\Bigr] |0 \rangle. 
\label{gsg} 
\end{eqnarray}
Altogether, the four-sublattice quantum antiferromagnet $|{\rm AF} \rangle$ can be characterized 
by a perfect N\'eel order of the Ising spins situated at the nodal sites of some loose-packed 2D lattice, while the mobile electrons delocalized over its decorating sites rest in the entangled 
state composed of two intrinsic antiferromagnetic states $c^{\dagger}_{k1, \uparrow} c^{\dagger}_{k2, \downarrow} |0 \rangle$, $c^{\dagger}_{k1, \downarrow} c^{\dagger}_{k2, \uparrow} |0 \rangle$, and, 
two non-magnetic ionic states $c^{\dagger}_{k1, \uparrow} c^{\dagger}_{k1, \downarrow} |0 \rangle$, 
$c^{\dagger}_{k2, \uparrow} c^{\dagger}_{k2, \downarrow} |0 \rangle$. The quantum entanglement 
of those four microstates arises from a virtual hopping process of the itinerant electrons, 
which is diagrammatically illustrated in Fig. \ref{fig2}. The respective probability distribution 
of the four entangled microstates is displayed in Fig. \ref{fig3} as a function of a relative 
strength of the hopping term.

Let us make a few comments on an origin of the remarkable four-sublattice quantum antiferromagnet emerging in $|{\rm AF} \rangle$. The hopping process of both itinerant electrons energetically favors their antiparallel spin alignment and this antiferromagnetic correlation is subsequently mediated through the Ising-type exchange interaction $J$ also on their two nearest-neighbor Ising spins 
(hence, the antiferromagnetic character of the effective interaction $R<0$ between the Ising spins). 
If a relative strength of the hopping term is sufficiently weak, the Ising spins prefer the ferromagnetic alignment with respect to their nearest-neighbor mobile electrons as it can 
be clearly seen from the occurrence probability of the majority microstate 
$c^{\dagger}_{k1, \uparrow} c^{\dagger}_{k2, \downarrow} |0 \rangle$ that converges 
to $p_{\uparrow,\downarrow} \to 1$ in the limit $t/J \to 0$. It is nevertheless worth mentioning 
that the occurrence probabilities of three minority microstates monotonically increase upon strengthening the hopping term at the expense of the occurrence probability of the majority microstate until they asymptotically reach the same value $p_{\uparrow,\downarrow} = p_{\downarrow,\uparrow} = p_{\uparrow \downarrow, 0} = p_{\downarrow \uparrow, 0} \to 1/4$ in the limit $t/J \to \infty$.

\subsection{Finite-temperature phase diagram} 

\begin{figure}[t]
\includegraphics[width=8cm]{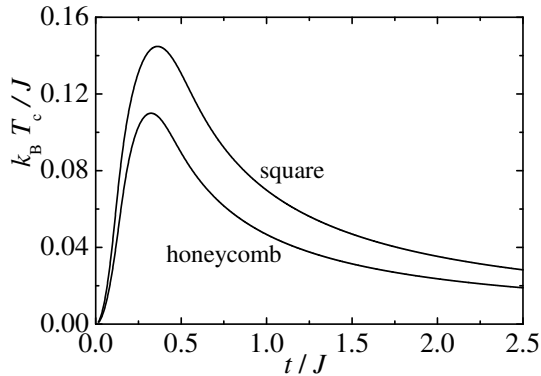}
\vspace{-0.5cm}
\caption{The dimensionless critical temperature as a function of the relative strength 
of the hopping term for the interacting spin-electron system on the doubly decorated 
square and honeycomb lattices.}
\label{fig4}
\end{figure}

Next, let us examine critical phenomena associated with finite-temperature phase transitions 
of the spontaneously long-range ordered phase $|{\rm AF} \rangle$. It is worthwhile to recall that 
the critical temperature can easily be obtained by solving numerically the critical conditions 
derived in the foregoing section. The finite-temperature phase diagram in the form of 
the critical temperature versus the kinetic term dependence is shown in Fig.~\ref{fig4} 
for the interacting spin-electron system on doubly decorated square and honeycomb lattices. 
It is quite obvious from this figure that the interacting spin-electron system on any 
loose-packed doubly decorated 2D lattice exhibits the same general trends in the relevant 
dependences of the critical temperature. The critical temperature initially exhibits 
a relatively rapid increase from zero temperature with increasing the ratio between 
the hopping term $t$ and the exchange constant $J$ 
until it achieves its maximum value. The critical temperature then gradually decreases 
upon further increase of a relative strength of the kinetic energy before it again tends 
to zero temperature in the other particular limit $t/J \to \infty$. In the limit $t/J \to 0$, 
the observed zero critical temperature can be attributed to a localization of the itinerant electrons at particular decorating sites. In fact, there does not exist any other interaction between the itinerant electrons within the tight-binding model described by the Hamiltonian (\ref{ham}) except 
the effective interaction originating from their virtual hopping process. If the hopping process 
of the itinerant electrons is ignored, the interacting spin-electron system then effectively splits into independent fragments each of them having one central Ising spin coupled to the $q$ localized 
electrons from its nearest-neighbor decorating sites. In the other particular limit $t/J \to \infty$, the zero critical temperature results from the same probabilities of the four entangled microstates 
of $|{\rm AF} \rangle$ as it has been already discussed by the ground-state analysis. Owing to 
this fact, the decorating sites occupied by the mobile electrons have non-magnetic character, 
which is compatible with a disappearance of the effective interaction $R=0$ between the localized 
Ising spins that occurs in the particular limit $t/J \to \infty$. 

\subsection{Order parameter} 

\begin{figure}[t]
\includegraphics[width=8cm]{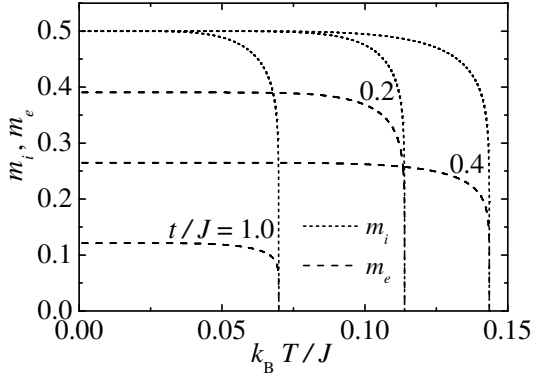}
\vspace{-0.5cm}
\caption{Thermal dependences of both sublattice staggered magnetizations for the interacting spin-electron system on the doubly decorated square lattice at three different values 
of a relative strength of the hopping term $t/J = 0.2$, $0.4$, and $1.0$.}
\label{fig5}
\end{figure}

Temperature variations of both sublattice staggered magnetizations $m_i$ and $m_e$ are depicted 
in Fig.~\ref{fig5} for three different values of a relative strength of the kinetic energy. 
The most obvious difference between the sublattice staggered magnetizations $m_i$ and $m_e$ 
pertinent to the Ising spins and itinerant electrons, respectively, lies in the quantum reduction 
of the latter staggered magnetization. The greater a relative strength of the hopping term is, 
the greater quantum reduction of the staggered magnetization $m_e$ can be observed in Fig.~\ref{fig5}
in concordance with the relevant ground-state prediction of the lowest-energy eigenstate (\ref{gs}), 
as well as, the zero-temperature limit of the expression (\ref{mag}) both yielding
\begin{eqnarray}
m_e (T=0) = \frac{1}{2} \frac{J}{\sqrt{J^2 + (4t)^2}}.  
\label{maggs}
\end{eqnarray}
On the other hand, the sublattice staggered magnetization $m_i$ pertinent to the Ising spins 
always starts from its maximum possible value that implies a perfect N\'eel long-range 
order of the localized Ising spins. For completeness, it is also worthy of notice that both 
sublattice staggered magnetizations tend to zero in a vicinity of the critical temperature 
with the critical exponent $\beta = 1/8$ from the standard Ising universality class.

\subsection{Specific heat} 

\begin{figure}[t]
\includegraphics[width=8cm]{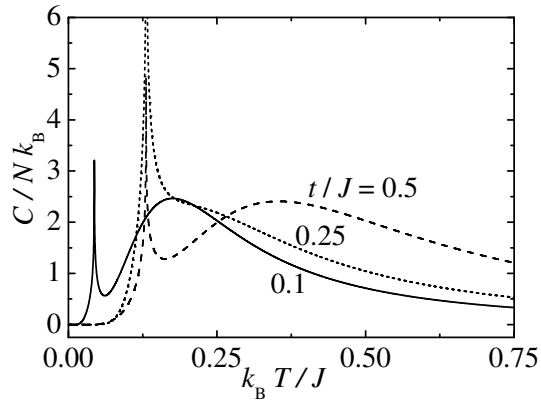}
\vspace{-0.5cm}
\caption{Temperature variations of the specific heat for the interacting spin-electron system 
on the doubly decorated square lattice at three different values of a relative strength of 
the hopping term $t/J = 0.1$, $0.25$, and $0.5$.}
\label{fig6}
\end{figure}

Finally, thermal dependences of the specific heat are plotted in Fig. \ref{fig6} for the interacting spin-electron system on the doubly decorated square lattice at three different relative strengths 
of the kinetic term $t/J = 0.1$, $0.25$, and $0.5$. It is quite evident that the investigated model system exhibits the familiar logarithmic singularity at critical temperature of the order-disorder transition, which provides another independent confirmation of the critical behavior from 
the standard Ising universality class. 
In addition to the logarithmic divergence observed in a close vicinity of the critical point, 
the specific heat also displays a round Schottky-type maximum in the high-temperature tail 
of the specific heat curve. This maximum is well separated from the logarithmic singularity 
at relatively weak hopping terms (see solid line for $t/J = 0.1$), while it becomes superimposed 
on a logarithmic divergence at moderate strengths of the kinetic term (see dotted line for 
$t/J = 0.25$). Last but not least, the round high-temperature maximum again separates from 
the logarithmic singularity upon further increase of a relative strength of the hopping term. 
The round maximum then shifts towards higher temperatures and it becomes the broader, the 
stronger a relative strength of the kinetic term is (see dashed line for $t/J = 0.5$).

\section{\label{sec:conc}Concluding remarks}

In this article, the hybrid lattice-statistical model of the interacting spin-electron system 
on doubly decorated 2D lattices has exactly been solved by the use of generalized decoration-iteration  transformation under the constraint of a half-filling of each couple of the decorating sites. 
It has been shown that the ground state of the investigated model system defined on any doubly decorated loose-packed 2D lattice represents an interesting four-sublattice quantum antiferromagnetic phase. It is worth noticing, moreover, that this four-sublattice quantum antiferromagnet exhibits 
a remarkable combination of the spontaneous long-range order manifested through a non-trivial criticality at non-zero temperatures with obvious macroscopic features of quantum origin such as 
the quantum reduction of the staggered magnetization pertinent to the itinerant electrons.
To the best of our knowledge, the model under investigation thus represents a rather rare example 
of the exactly solved model with such an intriguing combination of otherwise hardly compatible properties. To compare with, the interacting spin-electron system at a quarter filling (i.e. with one mobile electron per each couple of decorating sites) merely exhibits a classical ferromagnetic or ferrimagnetic spontaneous long-range order depending on whether the ferromagnetic or antiferromagnetic interaction between the localized Ising spins and itinerant electrons is assumed.\cite{icm} 
This would indicate that an existence of the four-sublattice quantum antiferromagnetic phase 
is closely related to a collective motion of electrons even if there does not exist any direct interaction between mobile electrons within our tight-binding model.  

Besides the purely academic interest in searching exactly solvable quantum spin models, the considered model system has also been suggested in order to bring insight into a magnetism of hybrid systems consisting of both localized spins as well as mobile electrons. Among the most famous magnetic materials, which obey this specific requirement, one could mention the magnetic metal
SrCo$_6$O$_{11}$ \cite{ishi05,muku06,ishi07a,ishi07b} or the series of polymeric coordination compounds [Ru$_2$(OOC$t$Bu)$_4$]$_3$[M(CN)$_6$] (M = Fe\cite{yosh02,miku06,vos05}, Cr\cite{vos05,mill05}).  
In the latter family of magnetic materials, one and three unpaired electrons of the trivalent metal ions such as Fe$^{3+}$ ($S = 1/2$) and Cr$^{3+}$ ($S = 3/2$) are localized at the corners of a
simple square lattice, whereas three unpaired electrons are delocalized over the mixed-valent dimeric unit Ru$_2^{5+}$ residing each bond of the square lattice.\cite{norm79} In this respect, the magnetic structure of this series of coordination compounds closely resembles the one of the suggested model 
system even if the electronic structure of the mixed-valent dimeric unit Ru$_2^{5+}$ would surely require more complex Hamiltonian in order to describe the double-exchange mechanism in the mixed-valent Ru$_2^{5+}$ dimeric unit. In this direction will continue our further work.

\begin{acknowledgments}
This work was supported by the Slovak Research and Development Agency under the contract 
LPP-0107-06. The financial support provided under the grants VEGA 1/2009/05 and VVGS 2/09-10 
is also gratefully acknowledged.
\end{acknowledgments}


\begin{thebibliography}{100}

\bibitem{baxt82} 
R.J. Baxter, {\it Exactly solved models in statistical mechanics} (Academic Press, New York, 1982). 

\bibitem{matt93} 
D.C. Mattis, {\it The Many-Body Problem: An Encyclopedia of Exactly Solved Models in One Dimension} 
(World Scientific, Singapore, 1993).

\bibitem{sach99} 
S. Sachdev, \textit{Quantum Phase Transitions} (Cambrige University Press, Cambridge, 1999).

\bibitem{lavi99} 
D.A. Lavis and G.M. Bell, {\it Statistical Mechanics of Lattice Systems}
(Springer, Berlin, 1999), Vol.1.

\bibitem{lieb04} 
E.H. Lieb, {\it Condensed Matter Physics and Exactly Soluble Models}, 
edited by B. Nachtergaele, J.P. Solovej, and J. Yngvason (Springer, Berlin, 2004).

\bibitem{suth04}
B. Sutherland, {\it Beautiful Models: 70 Years of Exactly Solved Quantum Many-Body Problems}
(World Scientific, Singapore, 2004).

\bibitem{diep04} 
H.T. Diep and H. Giacomini, in {\it Frustrated Spin Systems}, 
edited by H.T. Diep (World Scientific, Singapore, 2004).

\bibitem{wu09}
F.Y. Wu, \textit{Exactly Solved Models: A Journey in Statistical Mechanics} 
(World Scientific, Singapore, 2009).

\bibitem{fish59} 
M.E. Fisher, Phys. Rev. \textbf{113}, 969 (1959).

\bibitem{syoz72} 
I. Syozi, in {\it Phase Transitions and Critical Phenomena}, 
edited by C. Domb and M.S. Green (Academic Press, New York, 1972), Vol.1.

\bibitem{roja09} 
O. Rojas, J.S. Valverde, S.M. de Souza, Physica A \textbf{388}, 1419 (2009).

\bibitem{syoz51}
I. Syozi, Progr. Theor. Phys. {\bf 6}, 306 (1951).

\bibitem{onsa44} 
L. Onsager, Phys. Rev. {\bf 65}, 117 (1944).

\bibitem{gonc82}
L.L. Gon\c{c}alves, Physica A {\bf 110}, 339 (1982).

\bibitem{hori83}
T. Horiguchi and L.L. Gon\c{c}alves, Physica A {\bf 120}, 600 (1983).

\bibitem{gonc84}
L.L. Gon\c{c}alves and T. Horiguchi, Physica A {\bf 127}, 587 (1984). 

\bibitem{gonc86}
L.L. Gon\c{c}alves and T. Horiguchi, J. Phys. A: Math. Gen. {\bf 19}, 1449 (1986). 

\bibitem{sant86}
R.J.V. dos Santos, S. Coutinho, and J.R.L. de Almeida, J. Phys. A: Math. Gen. {\bf 19}, 3049 (1986).

\bibitem{sant95}
R.J.V. dos Santos and S. Coutinho, Int. J. Mod. Phys. B \textbf{9}, 3387 (1995).   

\bibitem{stre02}
J.~Stre\v{c}ka and M.~Ja\v{s}\v{c}ur, Phys. Rev. B \textbf{66}, 174415 (2002). 

\bibitem{jasc02}
J.~Stre\v{c}ka and M.~Ja\v{s}\v{c}ur, Phys. Stat. Solidi B \textbf{233}, R12 (2002). 

\bibitem{stre04}
J.~Stre\v{c}ka and M.~Ja\v{s}\v{c}ur, J. Magn. Magn. Mater. \textbf{272-276}, 987 (2004).

\bibitem{stre06}
J.~Stre\v{c}ka and M.~Ja\v{s}\v{c}ur, Acta Phys. Slovaca \textbf{56}, 65 (2006). 

\bibitem{stre08}
J.~Stre\v{c}ka, L.~\v{C}anov\'a, M.~Ja\v{s}\v{c}ur, and M.~Hagiwara, 
Phys. Rev. B \textbf{78}, 024427 (2008). 

\bibitem{yao08}
D.X.~Yao, Y.L.~Loh, E.W.~Carlson, and M.~Ma, Phys. Rev. B \textbf{78}, 024428 (2008). 

\bibitem{valv09}
J.S. Valverde, O. Rojas, and S.M. de Souza, Phys. Rev. E \textbf{79}, 041101 (2009). 

\bibitem{stre09}
J.~Stre\v{c}ka, L.~\v{C}anov\'a, and K.~Minami, Phys. Rev. E \textbf{79}, 051103 (2009). 

\bibitem{pere08} 
M.S.S. Pereira, F.A.B.F. de Moura, and M.L. Lyra, Phys. Rev. B \textbf{77}, 024402 (2008).

\bibitem{pere09} 
M.S.S. Pereira, F.A.B.F. de Moura, and M.L. Lyra, Phys. Rev. B \textbf{79}, 054427 (2009).

\bibitem{domb60}
C. Domb, Adv. Phys. \textbf{9}, 149 (1960).

\bibitem{mcoy73}
B.M. McCoy and T.T. Wu, \textit{The Two-Dimensional Ising Model} 
(Harvard University Press, Cambridge, 1973).

\bibitem{hout50}
R.M.F. Houtappel, Physica {\bf 16}, 425 (1950).

\bibitem{temp50} 
H.N.V. Temperley, Proc. Roy. Soc. A {\bf 203}, 202 (1950).

\bibitem{newe50}  
G.F. Newell, Phys. Rev. {\bf 79}, 876 (1950). 

\bibitem{husi50}   
K. Husimi and I. Syozi, Prog. Theor. Phys. {\bf 5}, 177 (1950).

\bibitem{syoz50} 
I. Syozi, Prog. Theor. Phys. {\bf 5}, 341 (1950).

\bibitem{wann50}
G.H. Wannier, Phys. Rev. {\bf 79}, 357 (1950); 
erratum: Phys. Rev. B \textbf{7}, 5017 (1973).
 
\bibitem{kano53}
K. Kano and S. Naya, Prog. Theor. Phys. {\bf 10}, 158 (1953).

\bibitem{barr88}
J.H. Barry, M. Khatun, and T. Tanaka, Phys. Rev. B {\bf 37}, 5193 (1988).

\bibitem{khat90}
M. Khatun, J.H. Barry, and T. Tanaka, Phys. Rev. B {\bf 42}, 4398 (1990).

\bibitem{barr91} 
J.H. Barry, T. Tanaka, M. Khatun, C.H. M\'unera, Phys. Rev. B {\bf 44}, 2595 (1991).

\bibitem{barr95}  
J.H. Barry and M. Khatun, Phys. Rev. B {\bf 51}, 5840 (1995).

\bibitem{acom}
Here and in what follows, the staggered magnetization is assumed to be calculated 
by adding a symmetry-breaking field to the Ising spin sites and taking the limit of zero field.  
 
\bibitem{lin92} 
K.Y. Lin, Chinese J. Phys. \textbf{30}, 287 (1992).

\bibitem{yang52} 
C.N. Yang, Phys. Rev. \textbf{85}, 808 (1952).

\bibitem{naya54} 
S. Naya, Prog. Theor. Phys. \textbf{11}, 53 (1954). 

\bibitem{call63} 
H.B. Callen, Phys. Lett. {\bf 4}, 161 (1963).

\bibitem{suzu65} 
M. Suzuki, Phys. Lett. {\bf 19}, 267 (1965).

\bibitem{saba81}
F.C. S\'aBarreto, I.P. Fittipaldi, and B. Zeks, Ferroelectrics \textbf{39}, 1103 (1981). 

\bibitem{saba85}
F.C. S\'aBarreto and I.P. Fittipaldi, Physica A \textbf{129}, 360 (1985).

\bibitem{balc02}
T. Balcerzak, J. Magn. Magn. Mater. \textbf{246}, 213 (2002). 

\bibitem{honm79} 
R. Honmura and T. Kaneyoshi, J. Phys. C: Solid St. Phys. {\bf 12}, 3979 (1979).

\bibitem{kane93}  
T. Kaneyoshi, Acta Phys. Pol. A {\bf 83}, 703 (1993).

\bibitem{note}
The relevant eigenstate for the other particular case $J<0$ can be obtained from the eigenvector (\ref{gsg}) under the reversal of all Ising spins $\sigma_{k}^z \to - \sigma_{k}^z$.  

\bibitem{icm} 
J. Stre\v{c}ka, A. Tanaka, M. Ja\v{s}\v{c}ur, presented at International Conference on Magnetism, 
to be held on July 27 -- 31, 2009, in Karlsruhe, Germany. Submitted to J. Phys.: Conf. Ser., 
preprint arxiv: 0908.2880. 

\bibitem{ishi05}
S. Ishiwata, D. Wang, T. Saito, and M. Takano, Chem. Mater. \textbf{17}, 2789 (2005).

\bibitem{muku06}
H. Mukuda, Y. Kitaoka, S. Ishiwata, T. Saito, Y. Shimakawa, H. Harima, and M. Takano,
J. Phys. Soc. Jpn. \textbf{75}, 094715 (2006). 

\bibitem{ishi07a}
S. Ishiwata, I. Terasaki, F. Ishii, N. Nagaosa, H. Mukuda, Y. Kitaoka, T. Saito, and M. Takano, 
Phys. Rev. Lett. \textbf{98}, 217201 (2007).

\bibitem{ishi07b}
S. Ishiwata, M. Lee, Y. Onose, N.P. Ong, M. Takano, and I. Terasaki, 
J. Magn. Magn. Mater. \textbf{310}, 1989 (2007).

\bibitem{yosh02}
D. Yoshioka, M. Mikuriya, and M. Handa, Chem. Lett., 1044 (2002).

\bibitem{miku06}
M. Mikuriya, D. Yoshioka, and M. Handa, Coord. Chem. Rev. \textbf{250}, 2194 (2006).

\bibitem{vos05}
T.E. Vos and J.S. Miller, Angew. Chem. Int. Ed. \textbf{44}, 2416 (2005).

\bibitem{mill05}
J.S. Miller, T.E. Vos, and W.W. Shum, Adv. Mater. \textbf{17}, 2251 (2005).

\bibitem{norm79}
J.G. Norman, G.E. Renzoni, and D.A. Case, J. Am. Chem. Soc. \textbf{101}, 5256 (1979).


\end{thebibliography}
\end{document}